\documentclass[11pt]{iopart}

\usepackage{graphicx}
\usepackage{iopams}
\usepackage{subfigure}
\usepackage{rotating}
\usepackage{color}

\begin{document}

\title{Resistive MHD Modelling of Quasi-Single Helicity State in the KTX Regimes}

{Bing Luo$^{a,b}$, Ping Zhu$^{a,b,c}$, Hong Li$^b$, Wandong Liu$^b$, and KTX team$^b$}

\address{$^a$CAS Key Laboratory of Geospace Environment and Department of Modern Physics, University of Science and Technology of China, Hefei, Anhui 230026, China}
\address{$^b$KTX Laboratory and Department of Modern Physics, University of Science and Technology of China, Hefei, Anhui 230026, China}
\address{$^c$University of Wisconsin-Madison, Madison, Wisconsin 53706, USA}
\ead{pzhu@ustc.edu.cn}
\vspace{10pt}
\begin{indented}
\item[]June 2017
\end{indented}

\begin{abstract}
 The potential formation of the quasi-single-helicity (QSH) state in the Keda Torus eXperiment (KTX) is investigated in resistive MHD simulations using the NIMROD code. We focus on the effects of finite resistivity on the mode structure and characteristics of the dominant linear and nonlinear resistive tearing-mode in a finite $\beta$, cylindrical configuration of reversed field pinch model for KTX. In the typical resistive regimes of KTX where Lundquist number $S=5 \times 10^4$, the plasma transitions to a steady QSH state after evolving through an initial transient phase with multiple helicities. The dominant mode of the QSH state develops from the dominant linear tearing mode instability. In lower $\beta$ regime, the QSH state are intermittent and short in duration; in higher $\beta$ regime, the QSH state persists for a longer time and should be more observable in experiment.
\end{abstract}
%
%
%
%

\section{Introduction}

The reversed-field pinch (RFP) is a toroidal magnetic confinement device. Its main difference from tokamak is its toroidal field $B_{\phi}$, which is of the same order of magnitude as the poloidal field $B_{\theta}$ and becomes reversed near plasma edge [1]. Since the safety factor $q(r)$ in RFP is always less than one, the RFP plasma can easily become unstable to resistive-kink modes.

During the last two decades, the QSH states have been discovered in four different RFP devices [2-5]: RFX in Padova [6], MST at the University of Wisconsin-Madison [7], EXTRAP T2R in Stockholm [8] and RELAX in Kyoto [9]. Significant efforts within the RFP community have been devoted to the study of quasi-single helicity (QSH) states ever since [10-12]. The main characteristic of the QSH state is the presence of an inner resonant dominant mode with poloidal periodicity $m=1$ and toroidal periodicity $n_{D}\ge {3R_{0}/2a}$. Its amplitude is typically several times larger than those of secondary modes. As a consequence, the QSH state plasma core has a helical deformation that generates a 3D structure like the stellarator [11]. QSH states with either Double Axis (DAx) or Single Helical Axis (SHAx) have been found in experiments [13,14]. The appearance of the QSH state reduces magnetic fluctuations and improves the RFP confinement, especially for the SHAx state, which has an inner electron temperature transport barrier [14,15]. The RFX-mod experimental statistics shows that the QSH state properties correlate with Lundquist number S, and higher plasma current leads to longer persistence of the QSH state [15-17]. Lately, the helical magnetic perturbation from the external active coils has been used to excite QSH with specific dominant modes in RFX-mod [18].

The proposition that RFP plasma could exist in a pure single helicity SH state was put forward in 1983, based on two-dimensional numerical simulations [19,20], The 2D simulations of pure (SH) states show the existence of two topologies for the corresponding magnetic surfaces: with or without a magnetic separatrix [12]. Later, numerical simulations of the RFP using 3D resistive magnetohydrodynamics codes SPECYl [21,22] also find MH and QSH states. Those simulations indicate the transition from MH to QSH state is controlled by the Hartmann number $H \propto \frac{1}{\sqrt{{\eta}{\nu}}}$ [23], where $\eta$ is resistivity, and $\nu$ is viscosity. If H is large, the system is in the MH state; when $H<3000$, the system displays temporal intermittency with a laminar phase of the QSH state [24,25]. The scheme of using the helical boundary condition generated by active coils to excite and control the specific dominate modes of QSH state in RFP has been studied using nonlinear 3D MHD simulations [18,26,27].

In this work, we use the 3D full MHD code NIMROD [28] to evaluate the possibilities of achieve QSH state in the newly constructed RFP device KTX, which is a middle-sized torus, with major radius $R_{0}=1.4$ m, minor radius $a=0.4$ m, and a designed maximum plasma current of 1 MA [29]. With 96 saddle coils installed outside of its copper shell, KTX is mainly designed to explore novel magnetic confinement regimes and advanced feedback control schemes for MHD instabilities, along with 3D physics. For the first time, our NIMROD simulations have demonstrated the potential formation of QSH state in the RFP configuration of KTX regimes. In particular, we find the connection between the QSH state and the dominant linear tearing mode on KTX. We further investigate the condition of QSH state formation and find a new parameter dependence of QSH, which is different from previous studies.

The rest of the paper is organized as follows. In Section 2, the single-fluid MHD model and a typical KTX equilibrium are described; in Section 3, the linear MHD instability of KTX is analysed including the growth rate and mode structure; in Section 4, the 3D QSH state is demonstrated in a nonlinear simulation, where a saturated magnetic island is observed to form and persist. Finally, Section 5 is devoted to a summary and discussion.

 \section{MHD model and equilibrium}
The numerical simulations in this work are performed with the single-fluid MHD model implemented in the NIMROD code. The single-fluid MHD equations can be written as follows:
 \begin{eqnarray}\label{eq1}
 ~\frac{\partial \rho}{\partial{t}}+\nabla \cdot(\rho \textbf{u})=0 \\
 ~\rho(\frac{\partial}{\partial{t}}+\textbf{v}\cdot \nabla)\textbf{v}=\textbf{J} \times \textbf{B}-\nabla p+\rho \nu \nabla^{2} \textbf{v}\\
 ~\frac{{N}}{(\gamma-1)}(\frac{\partial}{\partial{t}}+\textbf{v}\cdot \nabla)T=-p\nabla\cdot\textbf{v}-\nabla\cdot\textbf{q}\\
 ~\frac{\partial \textbf{B}}{\partial{t}}=\nabla \times (\textbf{v} \times \textbf{B}-\eta \textbf{J}) \\
 ~\nabla \times \textbf{B}=\mu_{0} \textbf{J} \\
 \textbf{q}=-N(\chi_{\parallel} \nabla_{\parallel} T+\chi_{\perp} \nabla_{\perp} T)
\end{eqnarray}
where $\rho$, $N$, \textbf{v}, $p$, \textbf{j}, \textbf{B}, \textbf{q}, $\gamma$, $\eta$, $\nu$, and $\chi_{\parallel}$ ($\chi_{\perp}$) are the plasma mass density, number density, velocity, pressure, current density, magnetic field, heat flux, specify heat ratio, resistivity, viscosity, and parallel (perpendicular) thermal conductivity, respectively.  The Lundquist number is defined as {$S$=$\tau_{R}$/$\tau_{A}$} with Alfven time $\tau_{A}$=$a$/$V_{A}$ ($V_{A}$=$B_{0}/{\sqrt{\mu_{0}\rho)}}$, resistivity diffusion time $\tau_{R}$=$\mu_{0} a^{2}$/$\eta$, and $B_0$ is the initial magnetic field strength on the magnetic axis.

In this work, we consider the low-$\beta$ regime of KTX plasmas, in contrast to the zero $\beta$ regime considered in the SPECY1 simulations [22,25]. Its initial configuration can be approximated as a cylindrical force-free RFP equilibrium, which can be specified using the following q profile (Fig. 1) [30]:
 \begin{equation}
  q(r)=q_0[1-1.8748(r)^2+0.83232(r)^4]\\
\end{equation}

 The above RFP equilibrium has been commonly adopted in previous studies [21,31,32]. For the KTX device, the equilibrium parameters at the magnetic axis are $q_{0}=2a/3R_{0}=0.18$, $\Theta_{0}=\mu_{0} a J_{\phi 0}/2B_{0}=1.46$, $B_{0}=1.0$ T, $N =1.2 \times 10^{19}$ $m^{-3}$, $J_{\phi_0}=6.15 \times 10^{3}$ $MA/m^{2}$, resistivity $\eta$ and viscosity $\nu$ share the same normalized radial profile $\frac{{\eta(r)}}{\eta_0}=\frac{{\nu(r)}}{\nu_0}$=${[1+(r/a)^{20}]}^2$. In our simulations, $3\times10^{-7}<\eta_0<3\times10^{-4}$, $1\times10^{4}<S<1\times10^{7}$, $\nu_0=1\times10^{-3}$, and $10^{3}<H<7\times10^{4}$. For an arbitrary initial perturbation, the MHD equations (1)-(6) are numerically solved in the cylindrical geometry, along with the periodic boundary conditions in the axial direction and the ideal wall boundary conditions in the radial direction.

\section{Linear MHD instability analysis}

For benchmark purpose, two kinds of grids for cylindrical geometry, namely ``circular" and ``rectangular", are been set up for the simulations (Fig. 2). Simulations results from the ``circular" and the ``rectangular" grids have been compared to verify their correctness. For the same equilibrium physical parameters, the time evolutions of the magnetic energy of the m=1, n=9 mode yield the same growth rate from those two grids. The mode structures from the two grids are also consistent with each other (Fig. 3). From now on, all results reported in this paper are obtained from simulations on the ``circular" grid.

The relationship between the growth rate $\gamma$ and the toroidal mode number $n$ shown in Fig.4a for different Lundquist number $S$, clearly indicates that the instability is located in plasma core, and the n=8 mode is the most unstable for this equilibrium. The normalized growth rate $\gamma \tau_{A}$ as a function of $S$ for the n=8 mode shown in Fig.4b, suggests that the instability is likely the resistive tearing mode in nature, since the asymptotic scaling $\gamma \tau_{A}$ $\propto$ $S^{-0.6}$ obtained from calculation agrees with the traditional theory [33,34].

The characteristic perturbations and the structure of instability are further analysed. Fig. 5 shows the radial profiles of the normalized perturbation $v_r$ and $b_r$ along the mid-plane in the cross section of the n=8 mode for $S=1\times10^{6}$. Here $v_r$ and $b_r$ are the radial components of the perturbed velocity and magnetic filed, respectively. The resonant surface locates at r=0.163 m, which is the location of the 1/8 resonant surface based on the q profile. Across that resonant surface, $v_r$ is anti-symmetric, and $b_r$ is symmetric, which is same as the parity of tearing-mode [35]. The 2D mode structure in the poloidal plane can be more clearly seen in Fig. 6, which indicates that the instability is dominated by the m=1 tearing mode.

\section{QSH state from nonlinear simulations}
For nonlinear simulations, we use the parameter ${\eta}_{0}=6.0 \times 10^{-5}$ $\Omega m$, ${\nu}_{0}=1 \times 10^{-3}$ $m^2/s$, density $N=1.2 \times 10^{19}$ $m^{-3}$, $B_{0}=1.0$ T, $H=1.5 \times 10^{3}$, $\beta=0.05$, and magnetic Lundquist number $S=5.0 \times 10^4$ on the equilibrium magnetic axis. The nonlinear evolution of magnetic energy is dominated by $n=7-12$ modes (Fig.7a). At t=0.35 ms, the instability mode evolves towards a saturated phase. During $t=0.35-1.1$ ms, the modes n=7, 8, and 9 compete strongly, but at t=1.1 ms, the energy of the perturbation begins to reside in the n=8 mode. After t=2.0 ms, with the decrease in the magnetic energy of the n=7 and 9 modes, the plasma relaxes to the QSH state, which is dominated by the n=8 mode. After t=5.6 ms, the plasma transitions from the QSH to the MH state. It is customary to describe the width of the toroidal spectrum of m=1 mode by using the $N_{s}$ parameter, which is defined as $N_{s}=[\sum_{n=1}(W_{1,n}/\sum_{n'=1}{W_{1,n'}})^2]^{-1}$, where $W_{1,n}$ is the magnetic energy of the (m=1, n) mode. A pure SH spectrum corresponds to $N_{s}=1$; for pure QSH state, $N_{s}$ may be less than 2. The time evolution of $N_{s}$ verifies that during those periods of QSH as indicated from magnetic energy evolution of each in modes(Fig. 7a), the $N_{s}$ number is indeed less than 2 (Fig. 7b).

The emergence of QSH can be further demonstrated in the time history of the Poincare plot in the poloidal plane(Fig.8). Here the plot at t=0 ms corresponds to the equilibrium state, t=1.5 ms the MH state, and the plot for $t=2.3-3.9$ ms correspond to a phase of the QSH state. A clearly shaped magnetic island appears and persists during the QSH state. Around t=5.6-6.2 ms, the island disappears and the plasma transitions from the QSH to the MH state. These Poincare plots show that the magnetic stochastic region is located in the plasma interior from $r=0.1$ to $0.2$ m, corresponding to the location of the $n=7-12$ inner modes. The size of the magnetic island is determined by the ratio of the secondary modes within a certain range. The amplitudes of the m=1 modes for different toroidal numbers are compared between the MH states at t=0.71 ms, 1.5 ms, 6.2 ms and the QSH states at t=3.1 ms, t=3.9 ms and t=5.16 ms (Fig.9), which again shows that the QSH state is dominated by the n=8 mode. Since the n=8 mode has the largest linear growth rate, the nonlinear simulations results suggest that the QSH state may have developed mainly from the dominant linear tearing mode.

The formation mechanism for QSH has long been a subject of theory and experimental studies. SPECY1 code simulation indicates that Hartmann number $H$ and helical magnetic perturbation can determine the QSH state [24-26]. Statistical results from RFX-mod experiment indicate that plasma current and Lundquist number $S$ correlate to the presence of the QSH state. In contrast, our simulation suggests that $\beta$ is the governing parameter for the formation of the QSH state in the KTX configuration and parameter regimes. The time dependence of the spectral spread $N_{s}$ for different $\beta$ shows clearly the correlation (Fig.10). For lower $\beta$, the QSH state appears intermittently in a shorter duration of time; for higher $\beta$, the QSH state persists in a longer duration of time. Therefore, it might be possible to obtain a QSH state through auxiliary heating of electrons even in regimes of low plasma current.

\section{Summary and discussion}
In summary, for the first time, NIMROD simulations find the formation of QSH state in KTX regimes with Lundqusit number $S=5 \times 10^4$. The simulation starts from a MH state, which transitions to a QSH state when secondary modes decrease in amplitude. The simulation indicates that the dominate mode of QSH state may develop from the linear mode that has the maximum growth rate. The plasma $\beta$ appears to be the key parameter that governs the emergence and duration of the QSH state. QSH form when $\beta$ is above certain threshold and a higher $\beta$ leads to a longer duration of QSH.

An effective method to achieve and maintain the QSH state may be the auxiliary heating that enhances the electron temperature, even in regimes of low plasma current. However, in the transition between the MH phase and the QSH phase, it is not clear what mechanism  suppresses the magnetic fluctuations of the secondary modes and maintain the dominate mode. In addition, how two-fluid effects and anomalous viscosity in high plasma current regimes may affect the QSH state should be also studied in future work.

\section{Acknowledgment}
This research was supported by the National Magnetic Confinement Fusion Program of China under Grant Nos. 2014GB124002 and 2015GB101004, the National Natural Science Foundation of China under Grant Nos. 11635008, the U.S. Department of Energy Grant Nos. DE-FG02-86ER53218 and DE-FC02-08ER54795, and the 100 Talent Program of the Chinese Academy of Science. We are grateful for the support from the NIMROD and KTX teams. This research used resources from the Supercomputing Center of University of Science and Technology of China, and the National Energy Research Scientific Computing Center, a DOE Office of Science User Facility supported by the Office of Science of the U.S. Department of Energy under Contract No. DE-AC02-05CH11231.

\section{Reference}
\numrefs{1}
\item Bodin H and Newton A. 1980 {\it Nucl. Fusion} {\bf 20} 1255
\item P. Martin. et al 2000 {\it Phys. Plasmas} {\bf 7} 1984
\item L. Marrelli. et al 2002 {\it Phys. Plasmas} {\bf 9} 2868
\item L. Frassinetti. et al 2007 {\it Phys. Plasmas} {\bf 14} 112510
\item R. IKEZOE. et al 2008 {\it Plasma Fusion Res.} {\bf 3} 029
\item G. Rostagni 1995 {\it Fusion Eng. Des} {\bf 25} 301
\item Dexter R.N. et al 1991 {\it Fusion Technol} {\bf 19} 131
\item Brunsell P.R. et al 2001 {\it Plasma Phys. Control. Fusion} {\bf 43} 1457
\item S. Masamune et al 2007 {\it J. Phys. Soc. Jpn} {\bf 76} 123501
\item D.F. Escande. et al 2000 {\it Plasma Phys. Control Fusion} {\bf 42} B243
\item D.F. Escande. et al 2000 {\it Phys. Rev. Lett} {\bf 85} 1662
\item D.F. Escande. et al 2000 {\it Phys. Rev. Lett} {\bf 85} 3169
\item R. Lorenzini. et al 2008 {\it Phys. Rev. Lett} {\bf 101} 025005
\item P. Piovesan. et al 2008 {\it Nucl. Fusion} {\bf 49} 085036
\item M. Gobbin. et al 2011 {\it Phys.Rev.Lett} {\bf 106} 025001
\item L. Carraro. et al 2009  {\it Nucl. Fusion} {\bf 49} 055009
\item R. Lorenzini. et al 2009 {\it Nature Phys.} {\bf 5} 570
\item D.Bonfiglio. et al 2013 {\it Phys. Rev. Lett} {\bf 111} 085002
\item E.J.Caramana. et al 1983 {\it Phys. Fluids} {\bf 26} 1305
\item D.D.Schnack. et al 1985 {\it Phys. Fluids} {\bf 28} 321
\item S.Cappello. et al 1992 {\it Phys. Fluids B} {\bf 4} 611
\item S.Cappello. et al 1996 {\it Nucl. Fusion} {\bf 36} 571
\item D. Montgomery. et al 1992 {\it Plasma Phys. Control. Fusion} {\bf 34} 1157
\item S.Cappello. et al 2000 {\it Phys. Rev. Lett} {\bf 85} 3838
\item S.Cappello. et al 2004 {\it Plasma Phys. Control. Fusion} {\bf 46} B313
\item M.Veranda et al 2013 {\it Plasma Phys. Control. Fusion} {\bf 55} 074015
\item D.Bonfiglio et al 2015 {\it Plasma Phys. Control. Fusion} {\bf 57} 044001
\item Sovinec et al 2004 {\it J. Comput. Phys.} {\bf 195} 355
\item Wandong Liu. et al 2014 {\it Plasma Phys. Control. Fusion} {\bf 56} 094009
\item D.D.Schnack et al 1987 {\it J. Comput. Phys} {\bf 70} 330
\item Finn J. M et al 1992 {\it Phys. Fluids B} {\bf 4} 1262
\item M.Onofri et al 2008 {\it Phys. Rev. Lett} {\bf 101} 255002
\item H.P. Furth et al 1963 {\it Phys. Fluids} {\bf 6} 459
\item H.P. Furth et al 1973 {\it Phys. Fluids} {\bf 16} 1054
\item J.P Freidberg et al 1981 {\it J. Plasma Phys.} {\bf 26} 177
\endnumrefs

\newpage
\begin{figure}[htbp]
  \centering
  \includegraphics[width=0.7\textwidth,clip]{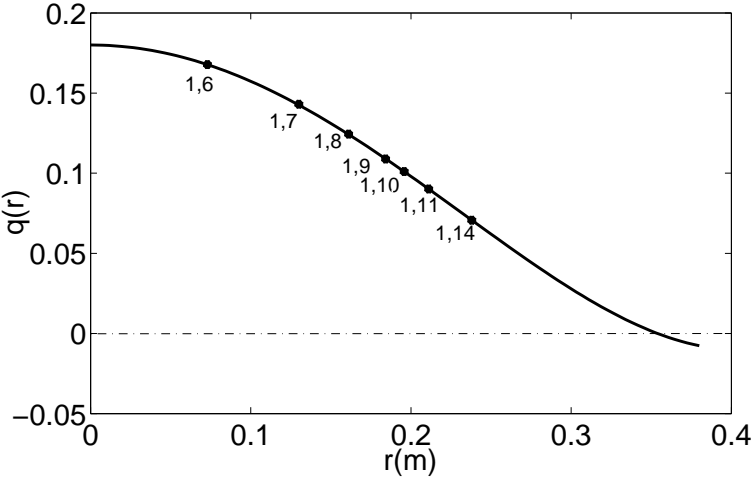}
  \caption{ Radial profile of the safety factor q(r) of the initial equilibrium for KTX.}\label{fig.1}
\end{figure}

\newpage
\begin{figure}[htbp]
  \centering
  \includegraphics[width=0.7\textwidth,clip]{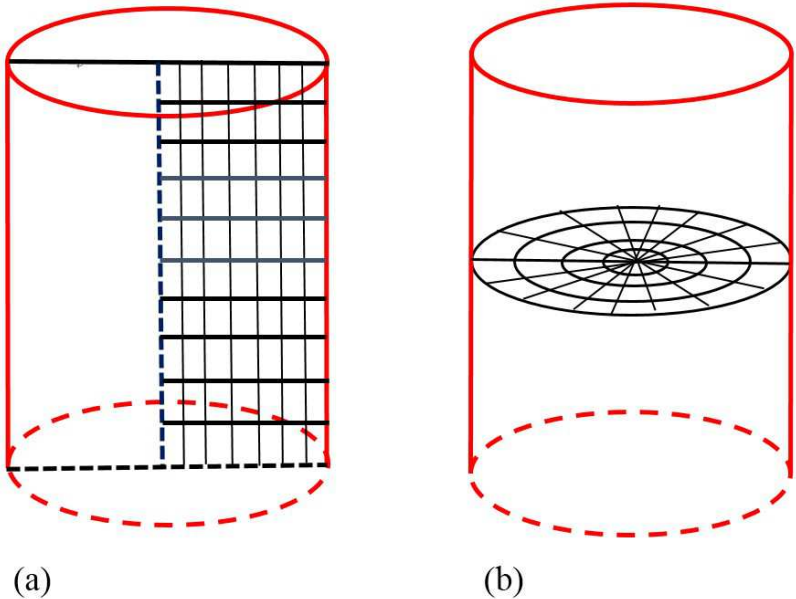}
  \caption{Two kinds of NIMROD grids for a cylindrical configuration: (left) ``rectangular" grid and (right) ``circular" grid.}\label{fig.2}
\end{figure}

\newpage
\begin{figure}[htbp]
  \centering
  \includegraphics[width=0.7\textwidth,clip]{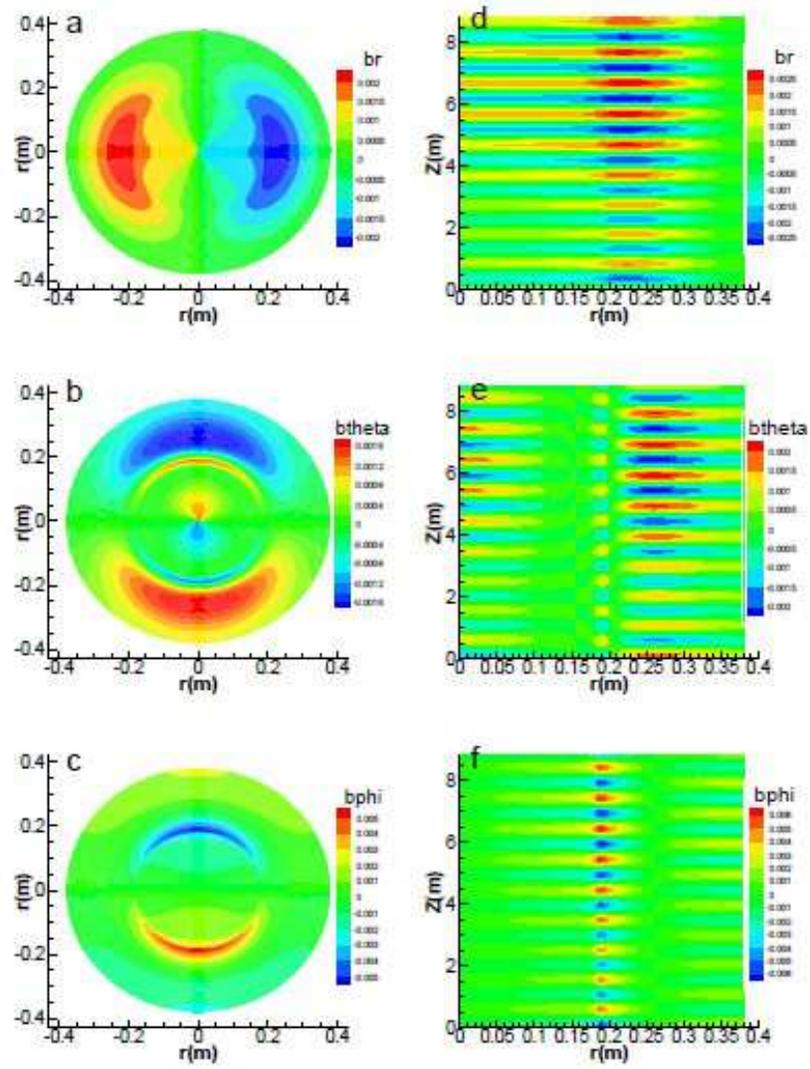}
  \caption{Contours of the radial, poloidal, and toroidal components of the perturbed magnetic field, i.e. $b_r$, $b_\theta$, and $b_\phi$ for the (m=1,n=9) mode in the $\phi=0$ plane of the ``circular" grid (a-c) and the $\theta=0$ plane of the  ``rectangular" grid (d-f).}\label{fig.3}
\end{figure}

\newpage
\begin{figure}[htbp]
  \centering
  \includegraphics[width=0.6\textwidth,clip]{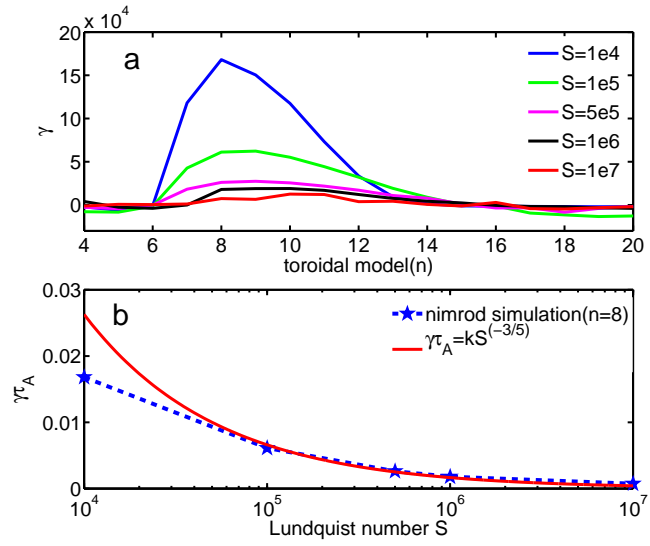}
  \caption{ (a) Linear growth rate as a function of toroidal mode number n for different values of Lundquist number S. (b) Linear growth rate as a function of Lundquist number S for the n=8 mode.}\label{fig.4}
\end{figure}

\newpage
\begin{figure}[htbp]
  \centering
  \includegraphics[width=0.6\textwidth,clip]{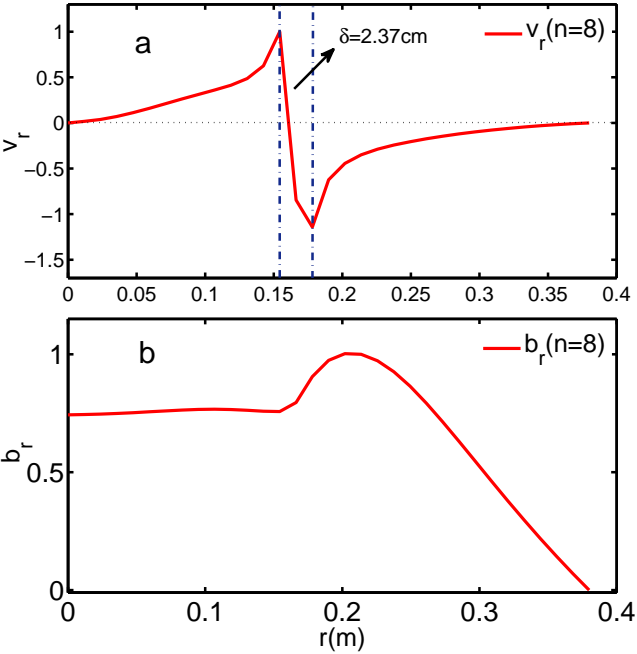}
  \caption{ Radial profiles as function of the minor radius r for (a) the normalized $v_{r}$ ($\theta=90^{o}$) and (b) the normalized $b_{r}$ ($\theta=180^{o}$) of the (m=1, n=8) mode with $S=10^{6}$. A singular layer locates around r = 0.163 m.}\label{fig.5}
\end{figure}

\newpage
\begin{figure}[htbp]
  \centering
  \includegraphics[width=0.37\textwidth,clip]{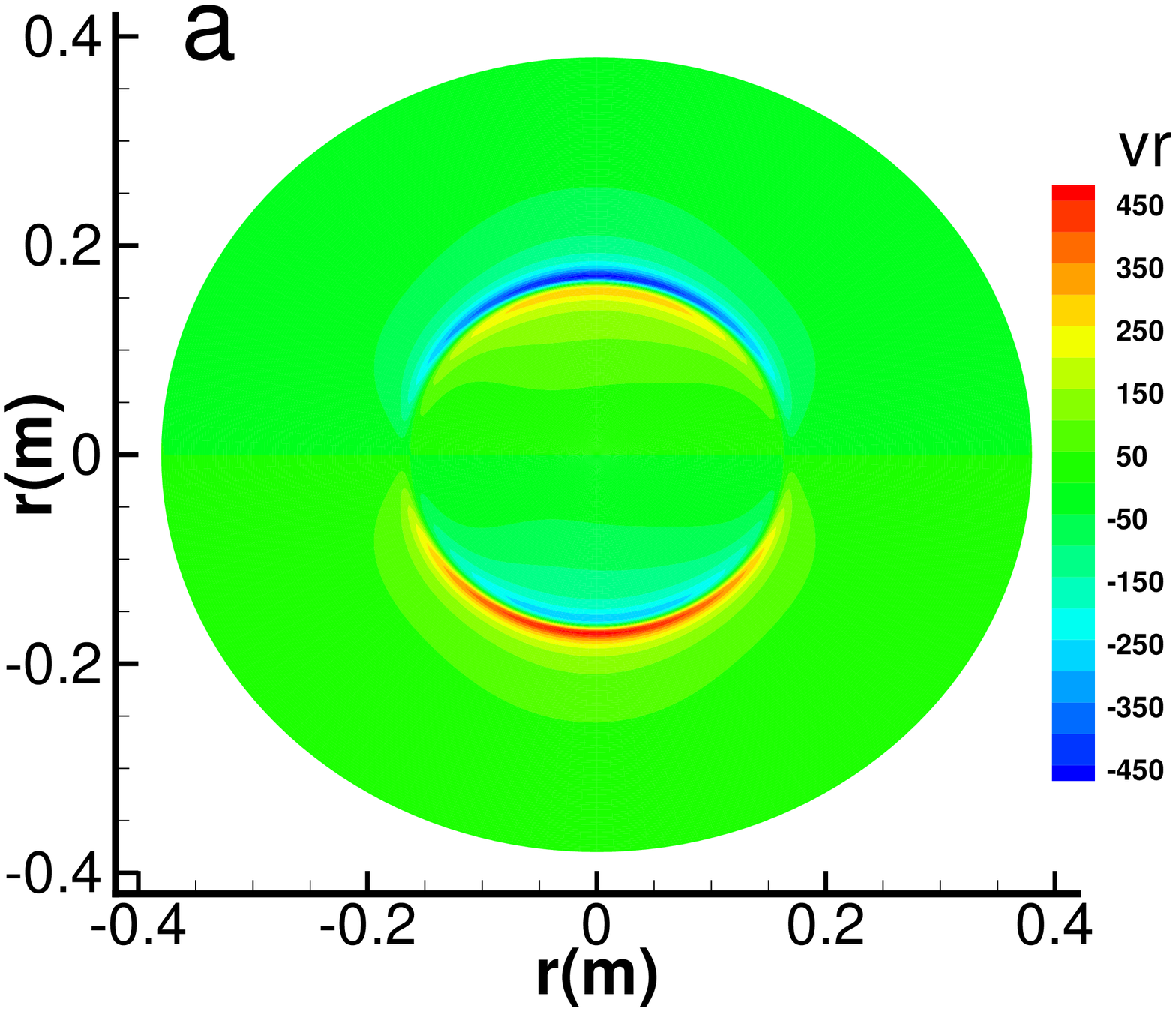}
  \includegraphics[width=0.37\textwidth,clip]{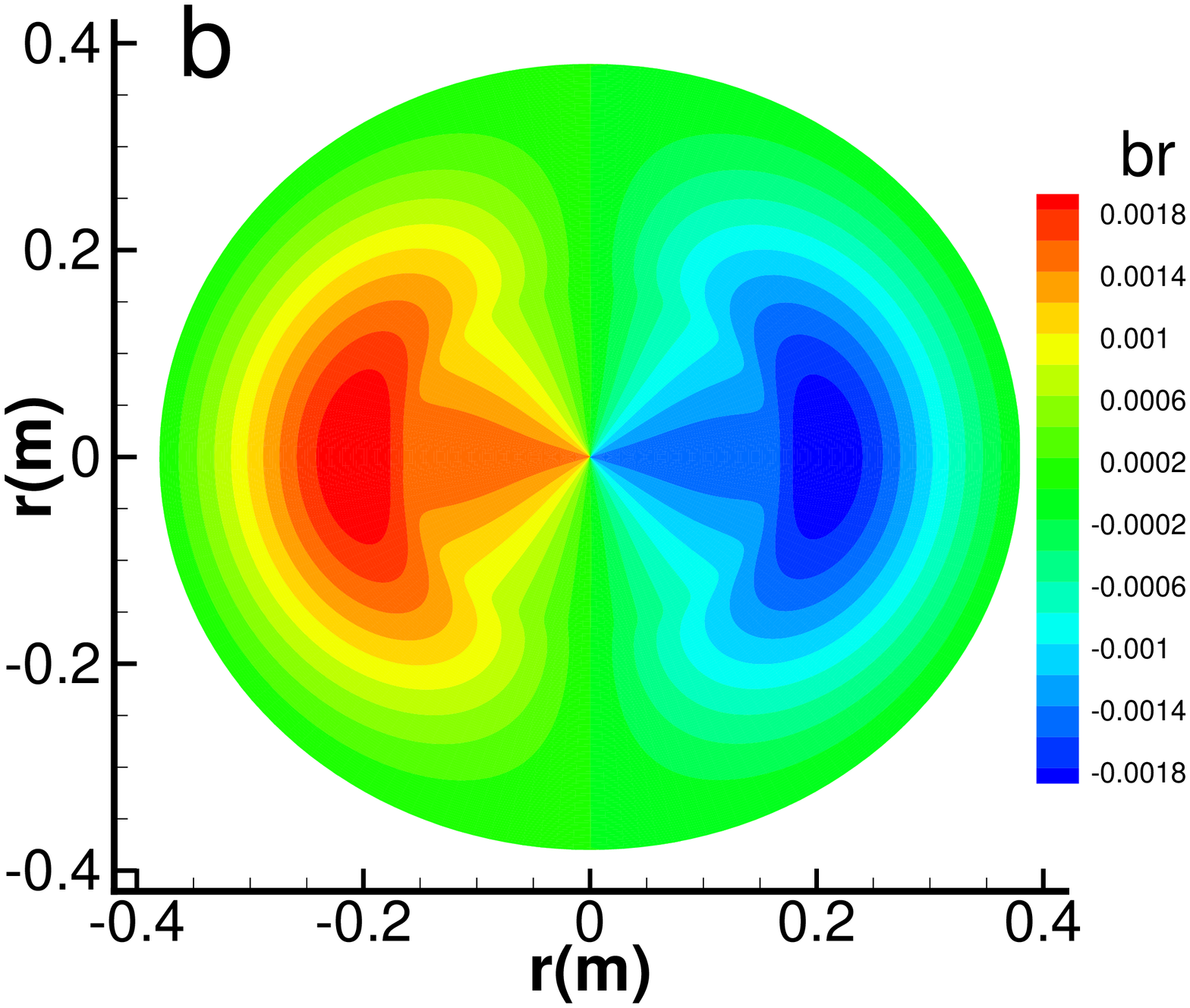}
  \caption{ Contours of (a) $v_{r}$ and (b) $b_{r}$ of  the (m=1, n=8) mode with $S=10^{6}$ in poloidal plane.}\label{fig.6}
\end{figure}

\newpage
\begin{figure}[htbp]
  \centering
  \includegraphics[width=0.75\textwidth,clip]{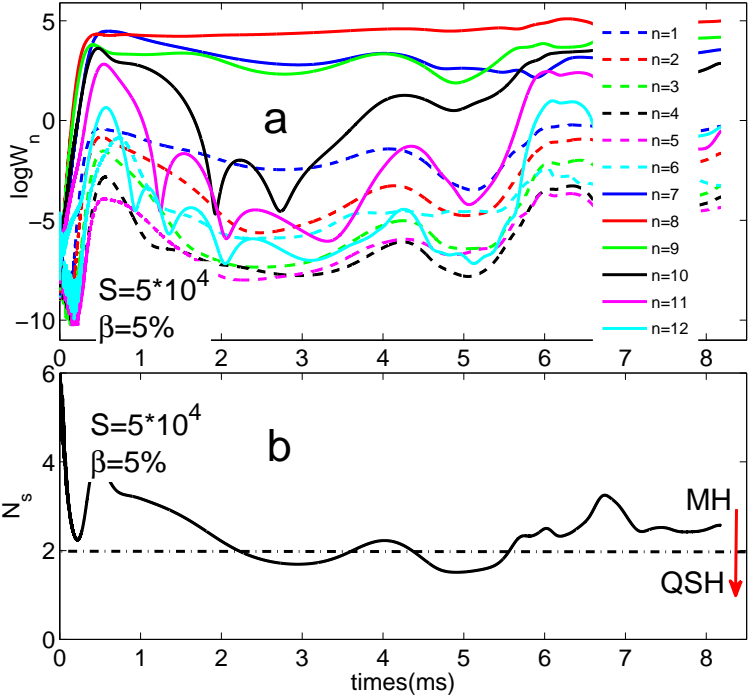}
  \caption{ (a) The magnetic energies of different toroidal mode components as functions of time, and (b) evolution of the spectral spread $N_{s}$ for $S=5.0 \times 10^4$ and $\beta=0.05$}\label{fig.7}
\end{figure}

\newpage
\begin{figure}[htbp]
  \centering
  \includegraphics[width=0.75\textwidth,clip]{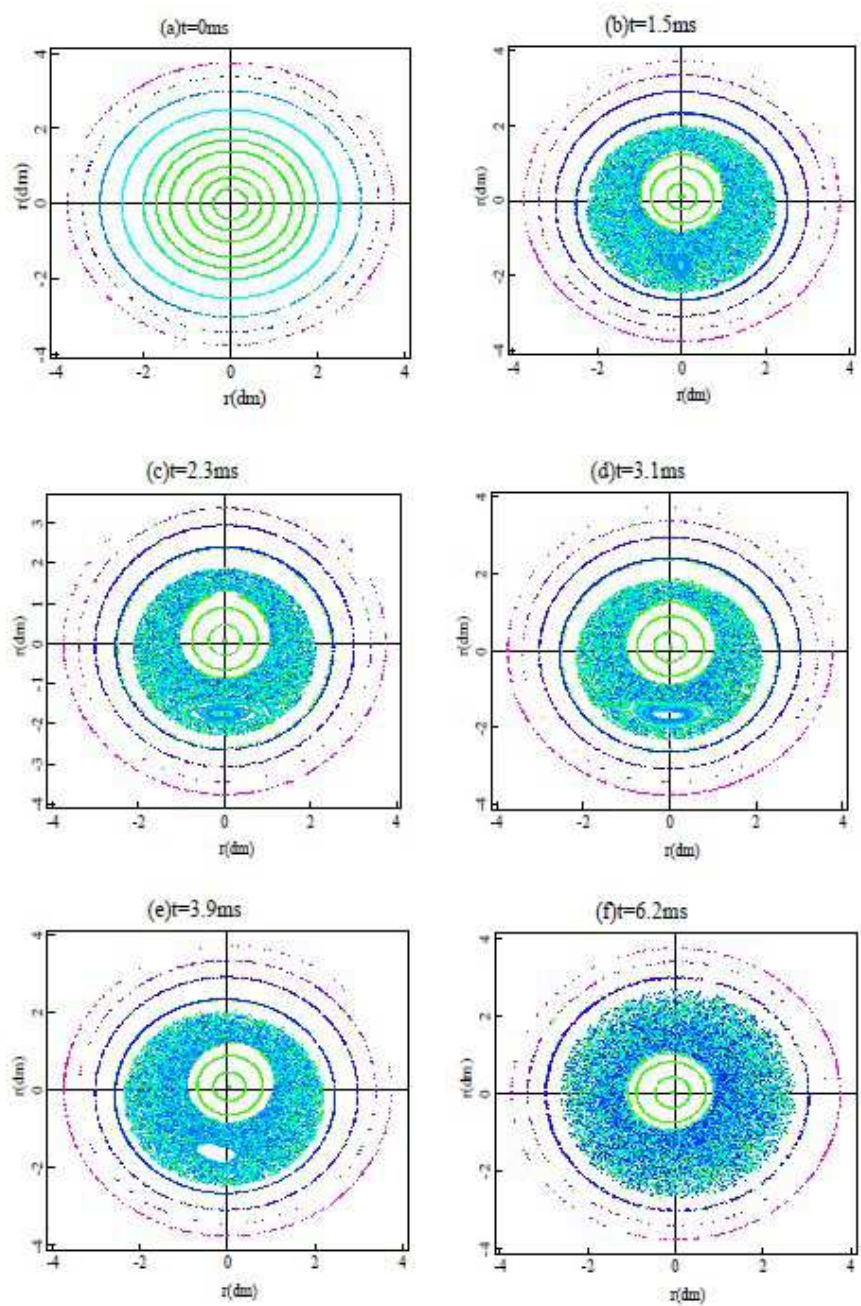}
  \caption{Snapshots of the Poincare plots in poloidal plane for the simulation case shown in Figure 7: $(a)-(f)$ correspond to t=0 ms, 1.5 ms, 2.3 ms, 3.1 ms, 3.9 ms, and 6.2 ms, respectively.}\label{fig.8}
\end{figure}

\newpage
\begin{figure}[htbp]
  \centering
  \includegraphics[width=9cm,height=9cm,clip]{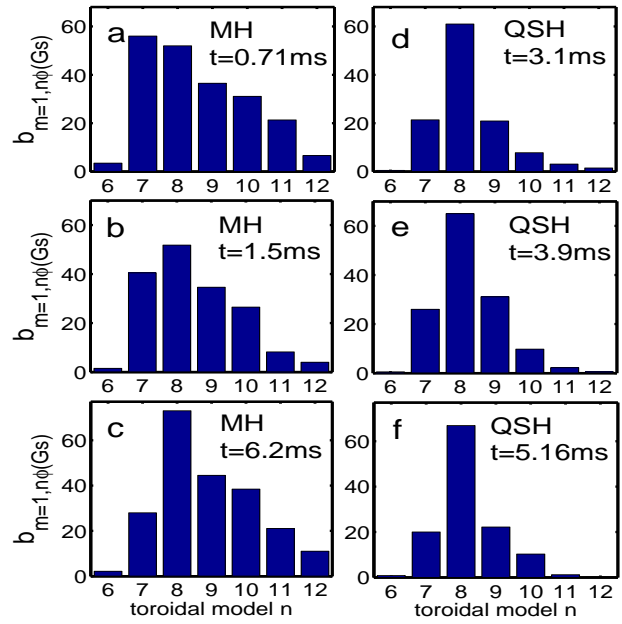}
  \caption{ The spectra in toroidal mode n for the m=1 mode at time t=0.71 ms, 1.5 ms and 6.2 ms in the MH state (a-c), and 3.1 ms, 3.9 ms, and 5.16 ms in the QSH state (d-f).}\label{fig.9}
\end{figure}

\newpage
\begin{figure}[htbp]
  \centering
  \includegraphics[width=11cm,height=7cm,clip]{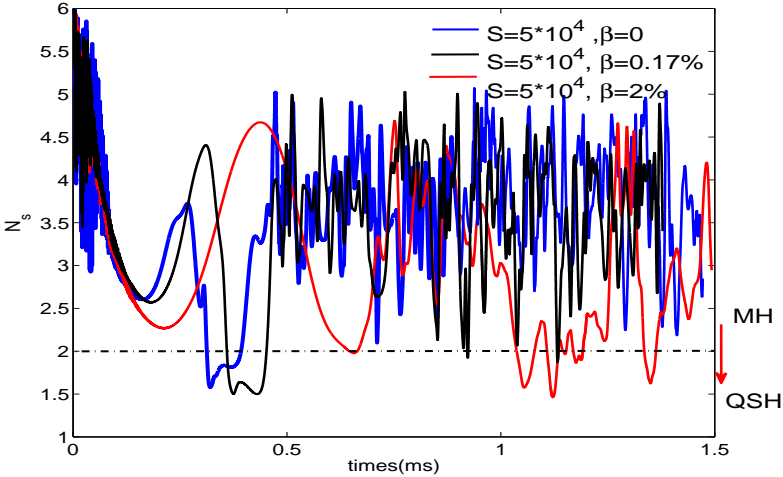}
  \caption{ Spectral spread in $N_{s}$ as a function of time for different $\beta$ with $S=5.0\times10^4$.}\label{fig.10}
\end{figure}

\end{document}